\documentstyle[psfig,epsfig,amssymb]{basi}
\begin{document}
\title[An H$\sc{i}$ study of long-tailed galaxies in Abell~1367]
      {An H$\sc{i}$ study of three long-tailed irregular galaxies in the cluster Abell~1367}
\author[Ananda Hota and D.J. Saikia]%
       {Ananda Hota$^{1,2,3}$\thanks{e-mail: hota@asiaa.sinica.edu.tw}
       and D.J. Saikia$^2$ \thanks{e-mail: djs@ncra.tifr.res.in} \\
   $^1$ Academia Sinica Institute of Astronomy and Astrophysics, P.O. Box 23-141, Taipei 10617, \\ Taiwan, Republic of China \\
   $^2$ National Centre for Radio Astrophysics (TIFR), Ganeshkhind, Pune 411 007, India \\
   $^3$ JAP Program, Indian Institute of Science, Bangalore 560 012, India }
\date{Received 26 March 2007; accepted 31 May 2007}
\maketitle
\label{firstpage}
\begin{abstract}
We present the results on the distribution and kinematics of
H$\sc{i}$ gas with higher sensitivity and in one case of higher spectral resolution
as well than reported earlier, of three irregular galaxies
CGCG 097073, 097079 and 097087 (UGC 06697) in the cluster Abell 1367.
These galaxies are known to exhibit long (50$-$75 kpc)
tails of radio continuum and optical emission lines (H$\alpha$) pointing
away from the cluster centre and arcs of starformation on the opposite sides
of the tails. These features as well as the H{\sc i} properties, with two
of the galaxies (CGCG 097073 and 097079) exhibiting sharper gradients in
H{\sc i} intensity on the side of the tails, are consistent with the
H{\sc i} gas being affected by the ram pressure of the intracluster medium.
However the H{\sc i} emission in all the three galaxies
extends to much smaller distances than the radio-continuum and H$\alpha$ tails,
and are possibly still bound to the parent galaxies. Approximately 20$-$30 per
cent of the H{\sc i} mass is seen to accumulate on the downstream side due to
the effects of ram pressure.
\end{abstract}


\begin{table*}
 \centering
 \begin{minipage}{140mm}
  \caption{Basic data on these three galaxies.$^a$}
  \begin{tabular}{@{}ccccccccc@{}}
\hline

{\small Galaxy}$^a$ &{\small RA$^b$ }& {\small Dec$^b$ }& Type$^c$ & a$\times$b$^d$  & V$_{sys}$$^e$  &D$^f$ &S$_{H{\sc i}}$$^g$ & S$^h$ \\ 

{\small CGCG} &  h m s  &  $^\circ$ $^{\prime}$ $^{\prime\prime}$  & &  $^\prime$ $\times$ $^\prime$  & km &  Mpc  & Jy  & mJy \\
              &         &                                          & &                                & /s &       & km/s  &     \\
\hline
097073 & 11 42 56.4  & +19 57 58 & SAcd, pec & 0.5$\times$0.5 & 7275$\pm$6 & 86   &  1.0    &  25    \\
097079 & 11 43 13.4  & +20 00 17 & Irr       & 0.5$\times$0.4 & 7000$\pm$9 & 86   &  0.8    &  15    \\
097087 & 11 43 49.1  & +19 58 06 & Im        & 1.9$\times$0.3 & 6725$\pm$2 & 86   &  3.5    &  60    \\
\hline
\end{tabular}\hfill\break
$a$ Taken from the NASA Extragalactic Database (NED), unless stated otherwise.  \hfill\break
$b$ Optical position of the galaxy from NED in J2000 co-ordinates.\hfill\break                                          
$c$ Morphological type.\hfill\break
$d$ Optical major and minor axes.\hfill\break
$e$ Heliocentric systemic velocity. \hfill\break
$f$ Assumed distance to the object from Gavazzi et al. (2001b). For this distance 1$^{\prime\prime}$=417 pc.\hfill\break
$g$ Total H{\sc i} line flux density taken from Arecibo measurements by Gavazzi (1989). \hfill\break
$h$ Total 1.4 GHz radio continuum flux density from the VLA D-array observations of Gavazzi (1989). \hfill\break
\end{minipage}
\label{basicdata}
\end{table*}

\section{Introduction}
Clusters of galaxies, which are the most massive gravitationally bound systems 
in the Universe, provide us with an opportunity to study the effects of the local 
environment on the structure, evolution and star formation history of its constituent 
galaxies (e.g. Gunn \& Gott 1972; Dressler 1980; Boselli \& Gavazzi 2006). Besides 
interactions with neighbouring galaxies, the hot intracluster medium (ICM) is 
also likely to play an important role in influencing the observed properties of the
galaxies. Simulations of stripping and compression of gas in spirals falling into a cluster 
by the ram pressure 
of the ICM have shown this to have a range of manifestations (Abadi, Moore \& Bower 1999; 
Quilis, Moore \& Bower 2000; Vollmer et al. 2001; Schulz \& Struck 2001; Bekki \& Couch 2003;
Roediger \& Br\"uggen 2006). For example, 
while the galactic interstellar medium (ISM) may be largely stripped in a high ICM density region 
leading to a suppression of star formation, the star formation rate may be enhanced in
a less dense ICM region where the ISM is only compressed rather than being completely 
stripped. In addition to these effects, clusters of galaxies
also provide us with important insights towards understanding the formation of 
large-scale structures in the Universe (e.g. West, Villumsen \& Dekel 1991; Katz \& White 1993). Many of 
these clusters have sub-structures, suggesting that these are dynamically young systems 
still in the process of formation. 

        The cluster Abell 1367 is an interesting nearby system at a distance of $\sim$86 Mpc where
three subgroups containing a number of star-forming galaxies are falling into the core
of the cluster (see Cortese et al. 2004). The cluster lies at the intersection of 
two filaments and has been suggested to be a prototype of a dynamically young cluster
(e.g. Cortese et al. 2004, 2006). Optical and radio observations of individual galaxies
by Gavazzi et al. (1995, 2001a,b) also suggest infall of galaxies into the cluster core.
Gavazzi \& Jaffe (1987) reported the discovery of extended tails of radio continuum emission 
associated with three irregular galaxies in the north-west region of A1367, namely CGCG 097073,
097079 and 097087 (UGC 06697). Tails of H$\alpha$ emission associated with the radio tails
have also been reported by Gavazzi et al. (1984; 2001a,b). X-ray observations of UGC 06697
also suggest that interaction between the  ISM and  ICM plays a major role in the observed
structures (Sun \& Vikhlinin 2005).
We have listed the basic properties of these three galaxies in Table 1. 
All the three galaxies have an asymmetric radio structure
with a `head' in the up-stream side roughly towards the cluster centre and a `tail'
on the opposite down-stream side with the size of the radio emission exceeding the 
size of the optical galaxy (e.g. Gavazzi 1978; Gavazzi \& Jaffe 1987). The galaxies CGCG 097073 
and 097079 also exhibit an arc
of H{\sc ii} regions suggesting star formation on their leading edges approximately
towards the cluster centre (Gavazzi et al. 1995, 2001a,b). These features are consistent
with the paradigm where ram pressure due to the ICM is significantly affecting the
observed properties of the galaxies. Observations of the atomic H{\sc i} 
gas are also consistent with the ram pressure paradigm. 
Gavazzi (1989) found the galaxies to be deficient in H{\sc i} compared with field galaxies
from observations with the Arecibo telescope.  The reported interferometric observations 
of H{\sc i} which only plotted the locations of the peaks of emission in a few channels showed the gas to be
displaced in the direction of the radio tails (Gavazzi 1989; Dickey \& Gavazzi 1991; hereafter
referred to as DG91), while
the molecular gas content of the galaxies appeared to be normal with the distribution
exhibiting no strong asymmetries (Boselli et al. 1994).

        In this paper we present the detailed distribution and velocity field of
H{\sc i} gas in all the three galaxies, CGCG 097073, 097079 and 097087 (UGC 06697),
with better sensitivity and in one case, CGCG 097079, with higher spectral resolution as well,
using archival Very Large Array (VLA) data with both the C- and D-configurations. These data
were also reduced with the objective of trying to detect any H{\sc i} gas from the long tails seen
at other wavelengths. In this context it is relevant to note that Oosterloo \& van Gorkom (2005)
have reported the detection of an H{\sc i} tail $\sim$110 kpc long which has been formed by
gas stripped from the galaxy NGC4388 by ram pressure. The extent of the tail suggests that
gas could remain neutral for $\sim$10$^8$ yr in the intracluster medium. A similar feature of 
$\sim$75 kpc in length 
near the galaxy NGC4438 has been reported by Hota, Saikia \& Irwin (2007), but it is possible
that this feature may also be of Galactic origin. Although the observations of 
CGCG 097073, 097079 and 097087 (UGC 06697) reported here do not reveal long H{\sc i} tails, they
reveal new features which we compare with observations at other wavelengths
and simulations of ram pressure stripping.

\section{Observations and data analysis}
The observing log for the observations is presented in Table \ref{obslog},
which is arranged as follows. Column 1: name of the telescope where we list
the configuration for the observations. The program code for the observations
in 1988 is AG264 (Principal investigator: J. Dickey, DG91) while for those in 
1999 it is AB900 (Principal investigator: B. Burke). 
Columns 2 and 3: dates of the observations and
the time, t, spent on the source in hours; 
column 4: the  channel separation in units of kHz and km s$^{-1}$; column 5:
the bandwidth of the observations in units of MHz and km s$^{-1}$. 

The observations were made
in the standard fashion, with each source observation interspersed
with observations of the phase calibrator.  The primary flux
density and bandpass calibrator was 3C286 whose flux density was estimated on the
Baars et al. (1977) scale using the 1999.2 VLA values.
The data analysis was done using the
Astronomical Image Processing System (AIPS) of the National Radio
Astronomy Observatory. The AIPS task UVLIN was used
for continuum subtraction and the multi-channel data were then CLEANed using IMAGR.

\begin{center}
\begin{table}
  \caption{H\,{\sc i} Observation log }     

  \begin{tabular}{l c r c c}
\hline
 Telescope       & Observation     & Time &  Channel separation & Band width  \\
 Array           & date            & hrs  &  kHz, km/s          & MHz, km/s   \\
 (1)             & (2)             & (3)  &  (4)                & (5)         \\
\hline
VLA-D            & 26 March 1999   & 9    &  98, 22    & 3.1, 650    \\
VLA-C            & 02 April 1988   & 8    &  391, 86   &12.5, 2600   \\
VLA-D            & 25 July 1988    & 3    &  391, 86   &12.5, 2600    \\
\hline
\end{tabular}
\label{obslog}
\end{table}
\end{center}

\begin{figure}[h]
\begin{center}
\resizebox{2.5in}{!}
{\includegraphics[angle=-90]{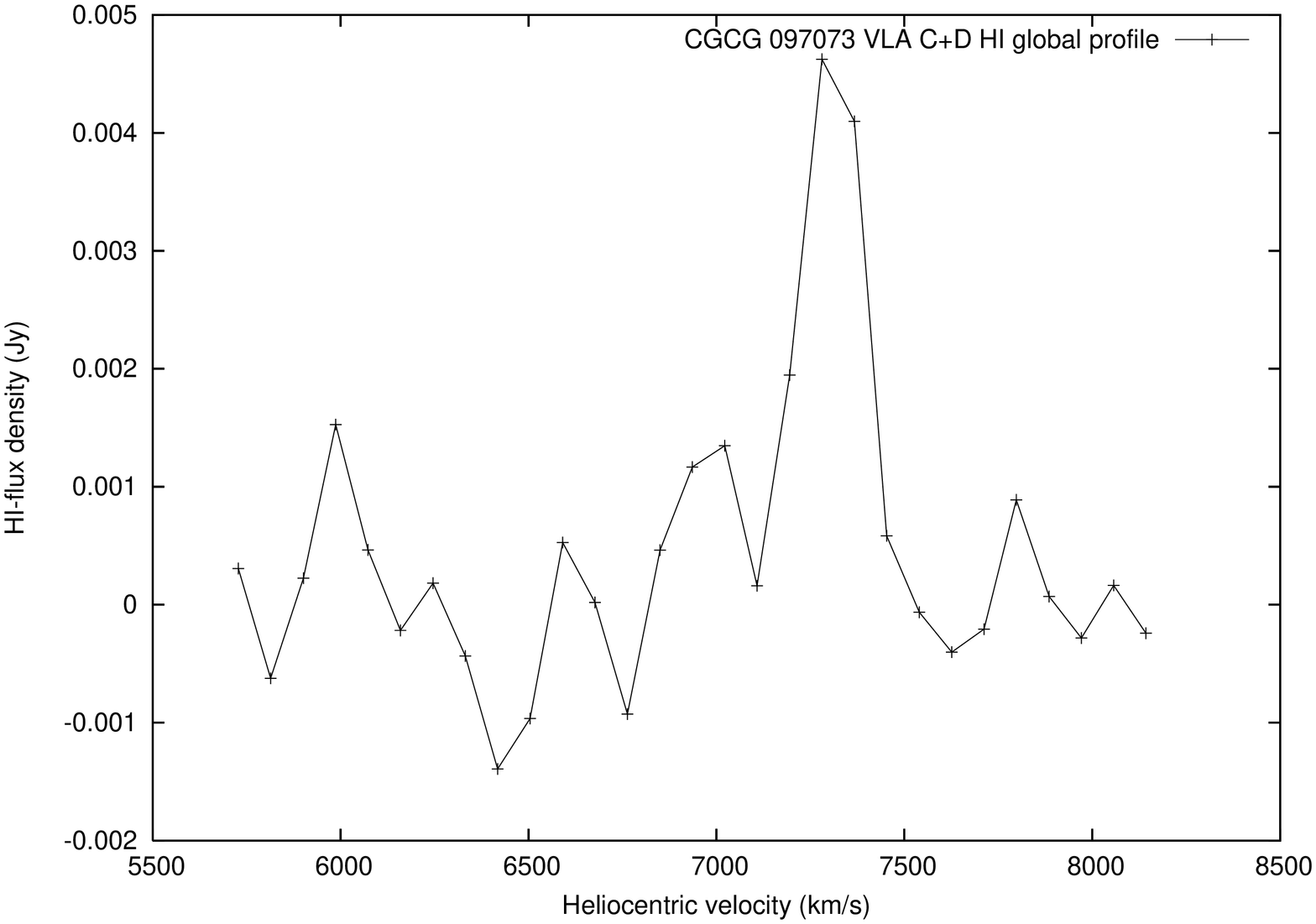}}
\caption{CGCG 097073: H{\sc i} global profile of the galaxy observed with a
spatial and spectral resolution of $\sim$21$^{\prime\prime}$ and 86 km s$^{-1}$ respectively.
}
\label{73.ispec}
\end{center}
\end{figure}

\begin{figure}[h]
\begin{center}
\resizebox{2.8in}{!} {\includegraphics[angle=0]{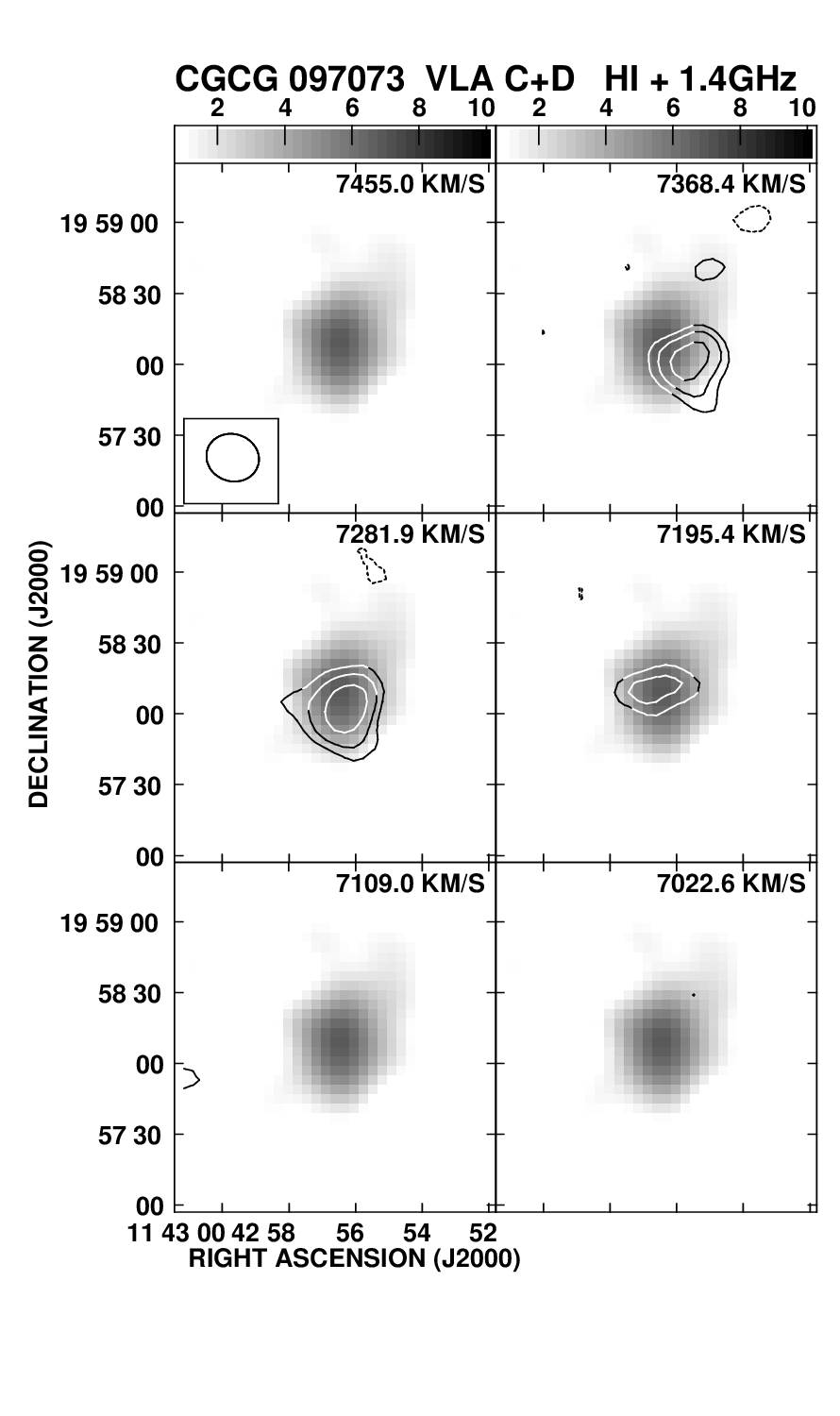}}
\vspace{-1.0cm}
\caption{CGCG 097073: The H{\sc i} channel maps in contours have been superimposed on the
gray scale single channel continuum image obtained from the same data with a resolution of $\sim$21$^{\prime\prime}$.
The contour levels are 0.3 $\times$ ($-$4, $-$2.82, 2.820, 4, 5.65 ...) mJy/beam. }
\label{73.chmap}
\end{center}
\end{figure}


\begin{table*}
  \caption{Observational parameters and some results from the H{\sc i} images}
  \begin{tabular}{lcccr rr ccc }
\hline
VLA     & Vel & \multicolumn{3}{c}{Beam size}                &  map  & spec.    & S$_{\rm H{\sc i}}$  & S$_{\rm H{\sc i}}$  & S$_{\rm H{\sc i}}$    \\
        & res.& maj.    & min.      & PA                         &  rms  & rms      & (097073)  & (097079)   & (097087)     \\
        &{\small km  }& $^{\prime\prime}$ & $^{\prime\prime}$ & $^\circ$ & {\small mJy} & mJy      &{\small Jy}                &  {\small Jy}       & {\small Jy}   \\
        &{\small   /s}&                   &                   &          & {\small /b} &          &{\small km/s}                 & {\small km/s}      &   {\small km/s}     \\
(1)     &(2)  &            (3) &     (4) &    (5) &  (6 ) &     (7 ) &    (8 )     &    (9 )      &    (10)          \\
\hline                                                                                                                                                                                 
D   &22   &    43.2            & 41.8             & $-$51.3    & 0.33  & 0.4      &                &  0.38           &                        \\
CD &86   &    22.5            & 19.9             & 71.8     & 0.30  & 0.7      &     0.92       &   0.52          &   2.67                \\
\hline

\end{tabular}\hfill\break
\label{obsparam}
\end{table*}

\section{Observational results}
The VLA C$-$ and D$-$array data which has a velocity resolution of 86 km s$^{-1}$
were combined to create a data cube with a spatial resolution of 
$\sim$21$^{\prime\prime}$ for all the three galaxies, while the VLA D-array data
with a velocity resolution of 22 km s$^{-1}$ was used to image the galaxy CGCG 097079.
The observational parameters and some results from the H{\sc i} images are presented
in Table \ref{obsparam} which is arranged as follows. Columns 1 and 2: the configuration of the
VLA observations and the spectral resolution in units of km s$^{-1}$; columns 3, 4 and
5: the major and minor axes of the restoring beam in arcsec and its position angle in
deg.; columns 6 and 7: the rms noise in the image and the spectrum in units of mJy/beam
and mJy respectively; columns 8, 9 and 10: the total H{\sc i} flux density in units of
Jy km s$^{-1}$ for the galaxies CGCG 097073, 097079 and 097087 respectively. 
 
\subsection{CGCG 097073}

\begin{figure}[!h]
\begin{center}
\resizebox{3.2in}{!}
{\includegraphics[angle=0]{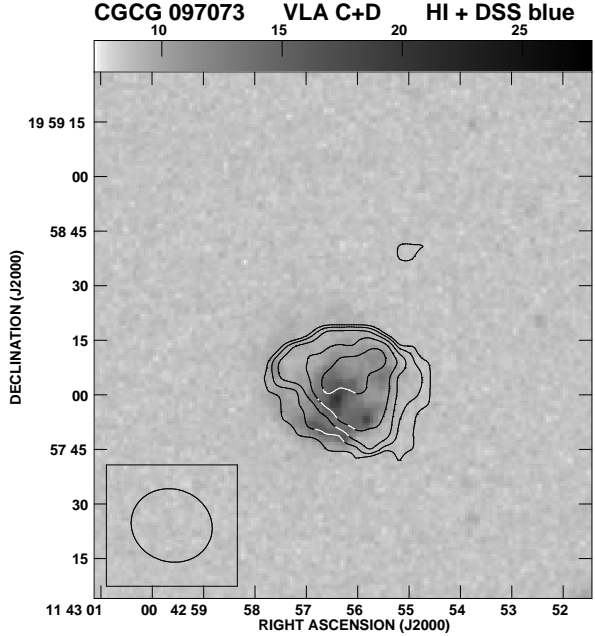}}
\vspace{-0.8cm}
\caption{CGCG 097073: Total intensity H{\sc i} contour map with a resolution of
$\sim$21$^{\prime\prime}$ has been superimposed on the DSS blue-band image.
The contour levels are 2.358 $\times$ 10$^{20}$ atoms cm$^{-2}$ or 1.89 M$_{\odot}$ pc$^{-2}$ and then
increasing in steps of $\sqrt{2}$.
}
\label{73.m0.dss}
\end{center}
\end{figure}

\begin{figure}[!h]
\begin{center}
\resizebox{2.2in}{!}
{\includegraphics[angle=0]{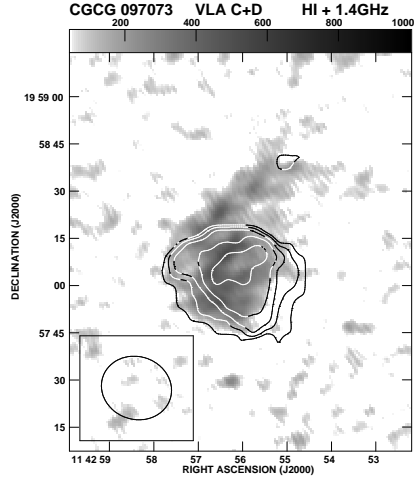}}
\vspace{-0.8cm}
\caption{CGCG 097073: Same H{\sc i} map on the 1.4-GHz radio continuum image made with a
resolution of $\sim$4$^{\prime\prime}$.
The contour levels are 2.358 $\times$ 10$^{20}$ atoms cm$^{-2}$ or 1.89 M$_\odot$ pc$^{-2}$ 
and then increasing in of $\sqrt{2}$.
}
\label{73.m0.radio}
\end{center}
\end{figure}

\begin{figure}[!h]
\begin{center}
\resizebox{2.2in}{!}
{\includegraphics[angle=0]{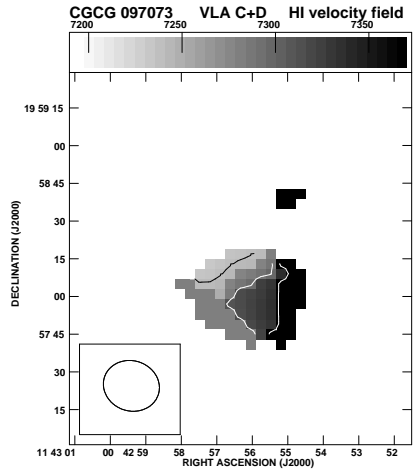}}
\vspace{-0.8cm}
\caption{CGCG 097073: The intensity weighted H{\sc i} velocity field
made from the same image cube with a spatial and spectral resolution of $\sim$21$^{\prime\prime}$
and 86 km s$^{-1}$ respectively. The contours are 7240, 7300 and 7360 km s$^{-1}$ from east to west.
}
\label{73.m1}
\end{center}
\end{figure}

The global profile obtained from the combined VLA C$-$ and D$-$array data with a spatial
resolution of $\sim$21$^{\prime\prime}$ and a velocity resolution of 86 km s$^{-1}$ 
does not show any significant asymmetry, consistent with the spectrum obtained with the Arecibo
telescope (Gavazzi 1989). Significant emission is seen in three channels whose velocities 
cover a range of $\sim$200 km s$^{-1}$. The width of the H{\sc i} spectrum obtained with the 
Arecibo telescope is 294 km s$^{-1}$. Both these values are significantly larger than the 
velocity width of 85 km s$^{-1}$ estimated from the Tully Fisher relation and inclination of 
the optical disk by Gavazzi (1989), suggesting strong kinematic effects leading to
non-circular motions in the ISM.
The total H{\sc i} mass estimated from the global profile is 1.6$\times$10$^9$
M$_\odot$. The peak of the H{\sc i} emission is consistent with the optical systemic 
velocity of 7225 km s$^{-1}$.  The H{\sc i} emission channel maps in
contours are shown superimposed on the radio-continuum, single channel gray scale image in Fig. \ref{73.chmap}.
It is clear that the position of the peak of H{\sc i} emission varies from channel to channel. 
DG91 reported H{\sc i} emission at 7282 and 7196 km s$^{-1}$ with the lower velocity detection towards
the northern edge of the optical image. We clearly see these two features but also emission
at 7368 km s$^{-1}$ towards the south-western part of the optical image. We have not been
able to confirm the possible weak component towards the north-west with a velocity of 7109 km s$^{-1}$
noted by DG91. 

\begin{figure}[h]
\begin{center}
\resizebox{2.2in}{!}
{\includegraphics[angle=-90]{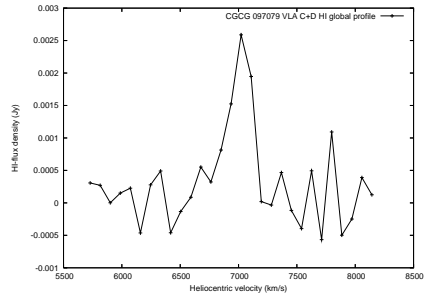}}
\caption{CGCG 097079: Global H{\sc i} profile of the galaxy made with a spatial and spectral
resolution of $\sim$21$^{\prime\prime}$ and 86 km s$^{-1}$ respectively.
}
\label{79.CD.ispec}
\end{center}
\end{figure}

\begin{figure}[h]
\begin{center}
\resizebox{2.2in}{!}
{\includegraphics[angle=-90]{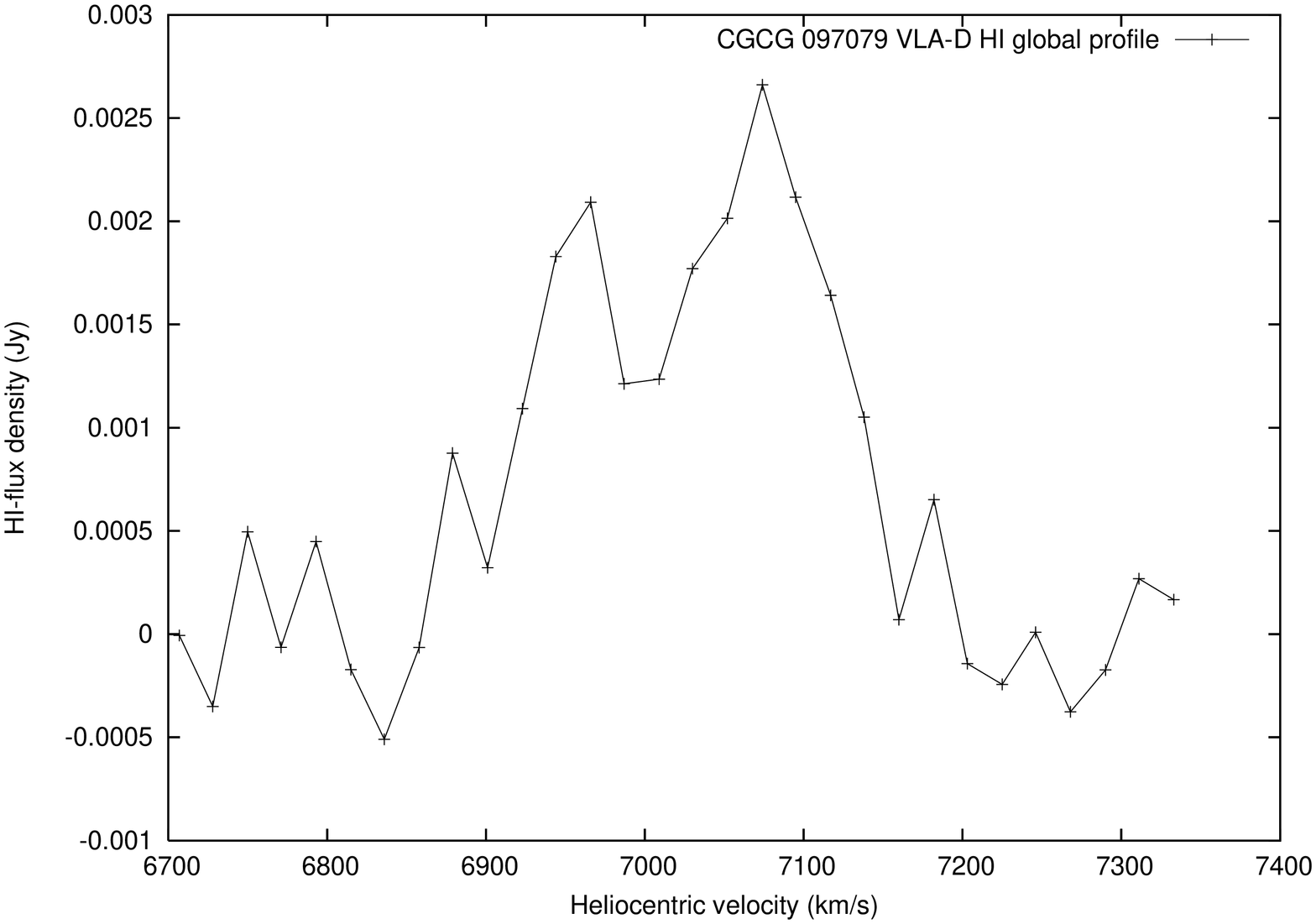}}
\caption{CGCG 097079: Global H{\sc i} profile of the same
galaxy made with a spatial and spectral resolution of $\sim$42$^{\prime\prime}$ and 22 km s$^{-1}$
respectively.
}
\label{79.D.ispec}
\end{center}
\end{figure}

\begin{figure}[h]
\begin{center}
\resizebox{2.2in}{!}
{\includegraphics[angle=-90]{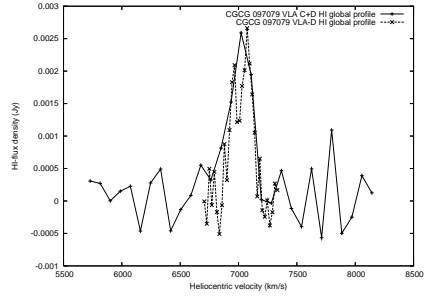}}
\caption{CGCG 097079: Global H{\sc i} profiles from the earlier two figures plotted together.
}
\label{79.added.ispec}
\end{center}
\end{figure}

\begin{figure}[!h]
\begin{center}
\resizebox{3.0in}{!}
{\includegraphics[angle=0]{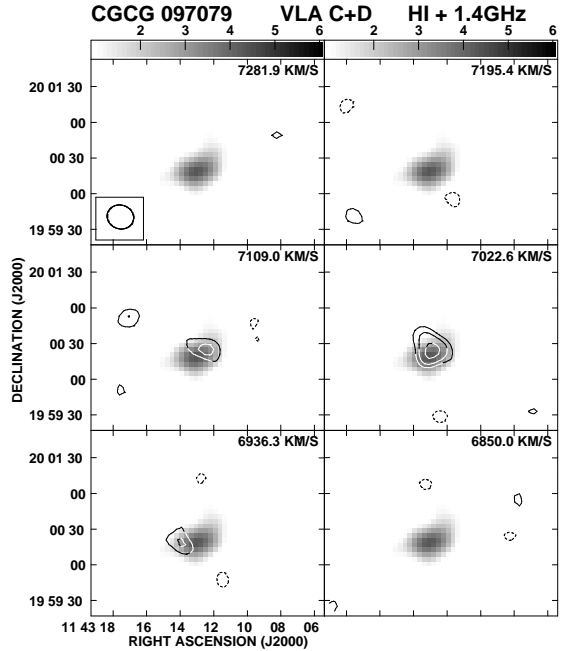}}
\vspace{-1cm}
\caption{CGCG 097079: The H{\sc i} channel maps in contours have been superimposed on the gray scale single channel
continuum image obtained from the same data with a resolution of $\sim$21$^{\prime\prime}$.
The contour levels are 0.3 $\times$ ($-$4, $-$2.82, 2.820, 4, 5.65 ...) mJy/beam.
}
\label{79.CD.chmap}
\end{center}
\end{figure}

A moment map generated from these three channels with a 3$\sigma$ cut off
is presented in Fig. \ref{73.m0.dss}. The total intensity H{\sc i}
(moment 0) map is shown superimposed on the DSS blue-band optical image of the galaxy. On the southern edge of 
the optical disk the stellar arc is visible with the H{\sc i} contours having a 
relatively sharper edge on the northern side.  This sharper
H{\sc i} contours on the opposite side of the optical starforming arc is also visible in the 
galaxy NGC2805, a  member of the group Ho 124 (Bosma et al. 1980; Kantharia et al. 2005). Coincidentally NGC2805 is 
also seen nearly face on and the galaxies in that group show evidence of ram pressure stripping and 
galaxy interactions. 
The accumulation of gas in the down-stream region can be qualitatively understood as being due to
the effect of ram pressure on the rotating gas. The rotating gas
following the ram pressure will reach the down-stream region faster while the gas rotating against
the ram pressure direction will face a greater resistance and thus spending a longer time
on the down-stream region.  Due to ram pressure one would also expect the up-stream side to get compressed 
and trigger star formation. Although at optical wavelengths there is an arc of H{\sc ii} regions
possibly triggered by compression of gas due to ram pressure, the H{\sc i} intensity contours do
not appear to be particularly edge-brightened.

The moment 0 image is also shown superimposed on a radio-continuum image at 1.4 GHz with an angular
resolution of $\sim$4$^{\prime\prime}$ made from archival VLA AB-array data (Fig. 4). At this resolution 
there is no clearly defined peak of radio continuum emission.
The southern arc-shaped region seen in radio continuum is also slightly offset from the southern
most peak of the arc seen at optical wavelengths.  Although this higher resolution image shows the 
orientation of the tail to the north-west, a larger-scale image with lower angular resolution shows that
the tail extends almost to the north (Gavazzi \& Jaffe 1987). We can see that on the northern side there is 
no correspondence of the H{\sc i} and radio continuum emission in the tail. Hence it is possible that we 
do not detect H{\sc i} from the stripped tail but the H{\sc i} gas is still largely rotating about 
the centre of the galaxy. Although the moment-one image shows evidence of rotation (Fig. 5), observations of
higher spectral resolution are required to determine the velocity field.  

\subsection{CGCG 097079}

\begin{figure}[!h]
\begin{center}
\resizebox{3.0in}{!}
{\includegraphics[angle=0]{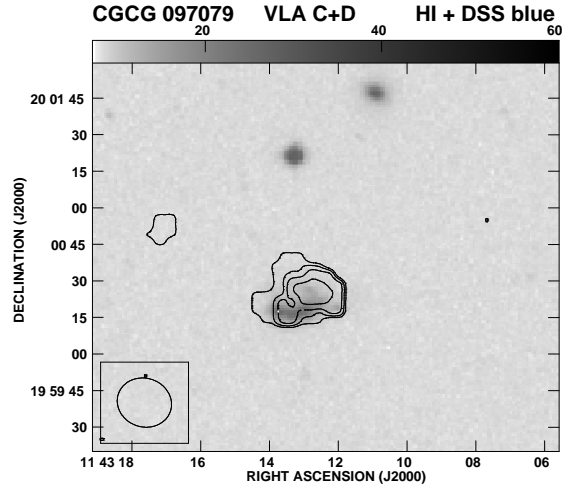}}
\caption{CGCG 097079: Total intensity H{\sc i} contour map with a resolution of $\sim$21$^{\prime\prime}$ has
been superimposed on the DSS blue-band image.  The contour levels are 2.358 $\times$ 10$^{20}$ atoms cm$^{-2}$ 
or 1.89 M$_\odot$ pc$^{-2}$ and then increasing in steps of $\sqrt{2}$.
}
\label{79.CD.MO.DSS}
\end{center}
\end{figure}

\begin{figure}[h]
\begin{center}
\resizebox{2.2in}{!}
{\includegraphics[angle=-90]{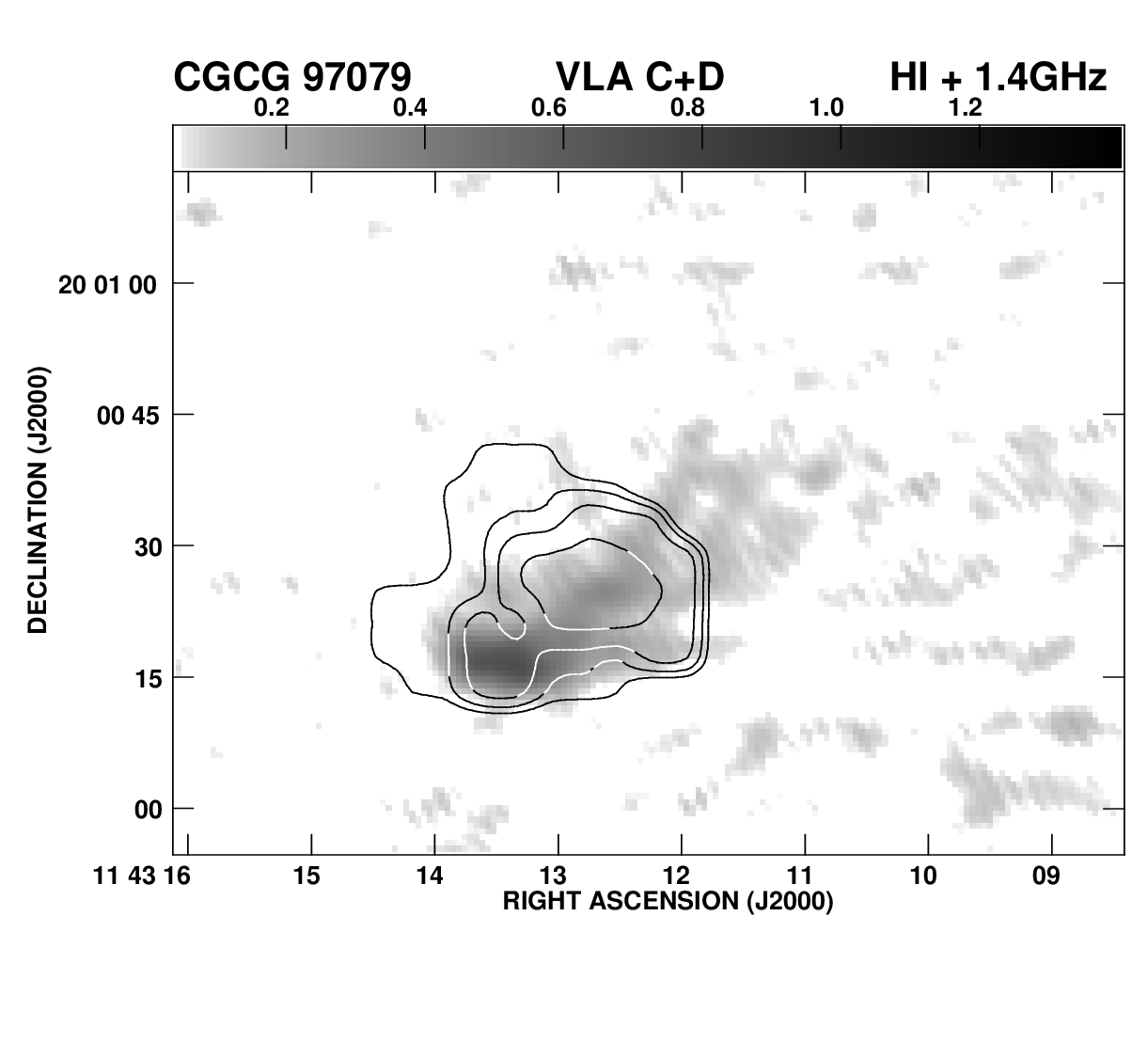}}
\caption{CGCG 097079: The same H{\sc i} map superimposed on the 1.4-GHz radio continuum image made with a
resolution of $\sim$4$^{\prime\prime}$. The contour levels are 2.358$\times$10$^{20}$ atoms cm$^{-2}$ or 
1.89 M$_\odot$ pc$^{-2}$ and then increasing in steps of $\sqrt{2}$.
}
\label{79.CD.M0.radio}
\end{center}
\end{figure}

\begin{figure}[h]
\begin{center}
\resizebox{2.2in}{!}
{\includegraphics[angle=-90]{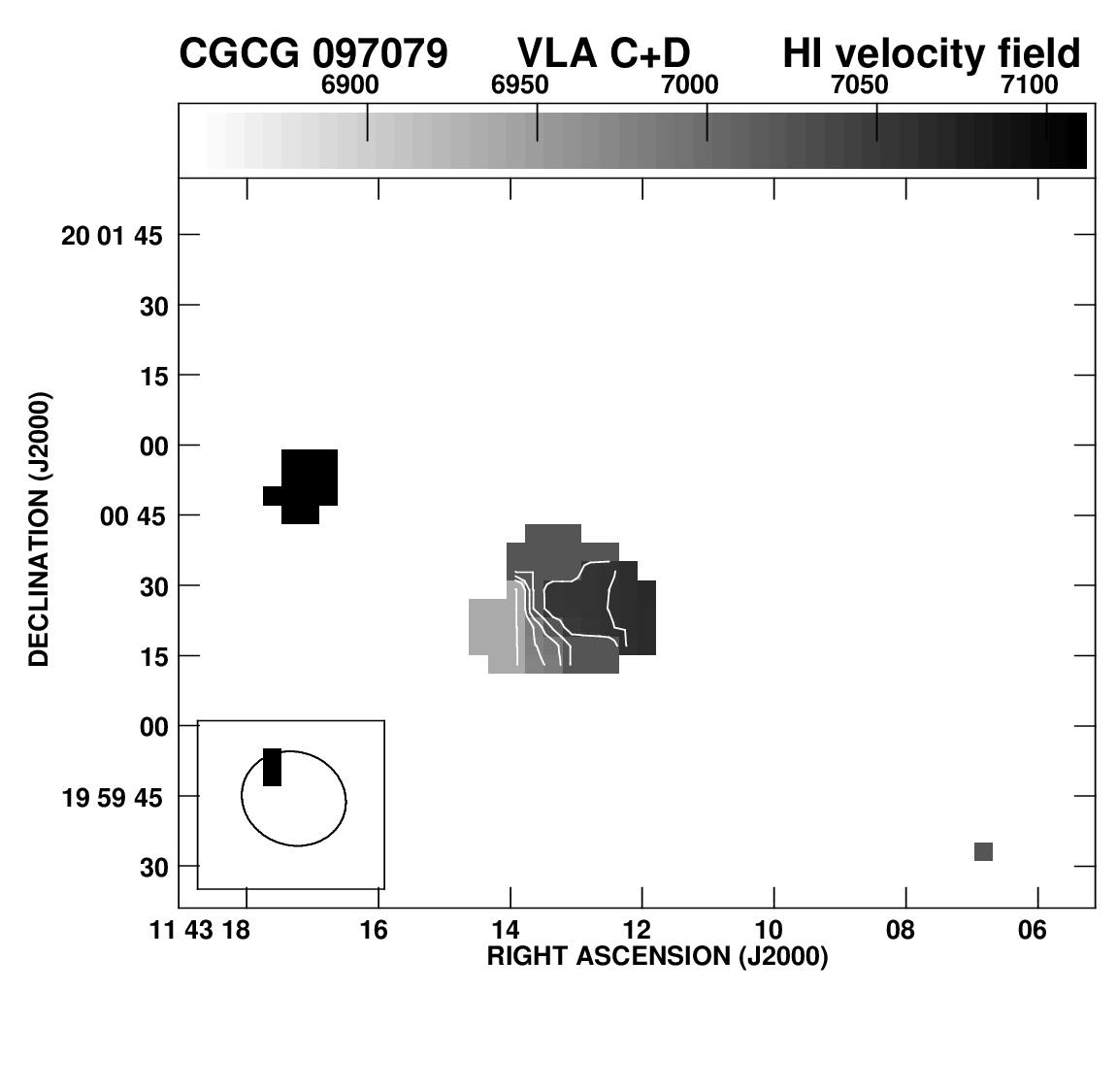}}
\caption{CGCG 097079: The intensity weighted H{\sc i} velocity field made from the
same image cube with a spatial and spectral resolution of $\sim$21$^{\prime\prime}$
and 86 km s$^{-1}$ respectively. The contours are 6940, 6980, 7000, 7020, 7040 and 7060 km s$^{-1}$ from east to west
}
\label{79.CD.M1}
\end{center}
\end{figure}

\begin{figure}[h]
\begin{center}
\resizebox{3.5in}{!}
{\includegraphics[angle=0]{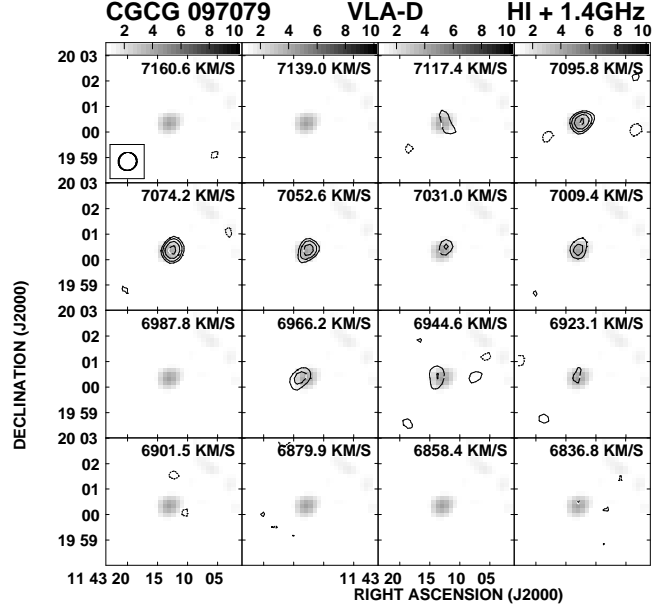}}
\vspace{-1cm}
\caption{CGCG 097079: The H{\sc i} channel maps in contours have been superimposed on the gray scale single-channel
continuum image obtained from the same data with a spatial and spectral resolution of $\sim$42$^{\prime\prime}$
and $\sim$22 km s$^{-1}$ respectively. The contour levels are 0.33$\times$($-$4, $-$2.82, 2.820, 4, 5.65 ...) mJy/beam.
}
\label{79.D.chmap}
\end{center}
\end{figure}

\begin{figure}[h]
\begin{center}
\resizebox{4.0in}{!}
{\includegraphics[angle=0]{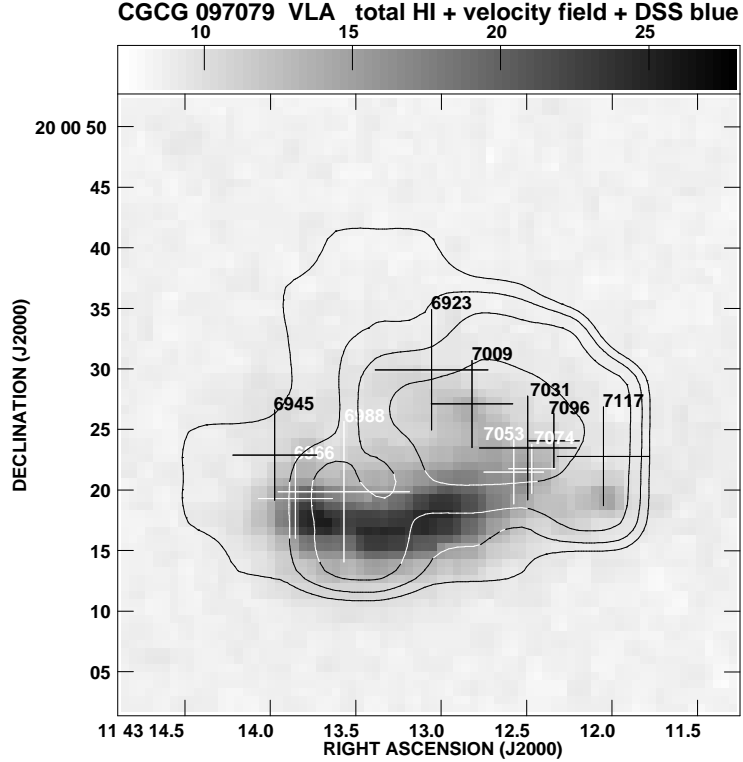}}
\vspace{-1cm}
\caption{CGCG 097079: The higher spatial resolution total intensity H{\sc i} contour of Fig. \ref{79.CD.MO.DSS}  
has been
superimposed on the DSS blue band image in gray scale. The `+' in the image marks the position of peak of the
Gaussian fit to the higher spectral resolution single channel H{\sc i} emission (Fig. \ref{79.D.chmap}). 
The size of the
'+' gives the uncertainty in defining the peak. The corresponding heliocentric velocity of the higher spectral
resolution channel emission has been marked close to the `+' mark. The contour levels are
2.358$\times$10$^{20}$ atoms cm$^{-2}$ or 1.89 M$_{\odot}$ pc$^{-2}$ and then increasing in steps of $\sqrt{2}$.
The diagonal velocity gradient and both the peaks of the H{\sc i} emissions are clearly visible.
}
\label{79.skull}
\end{center}
\end{figure}

\begin{figure}[!h]
\begin{center}
\resizebox{2.2in}{!}
{\includegraphics[angle=-90]{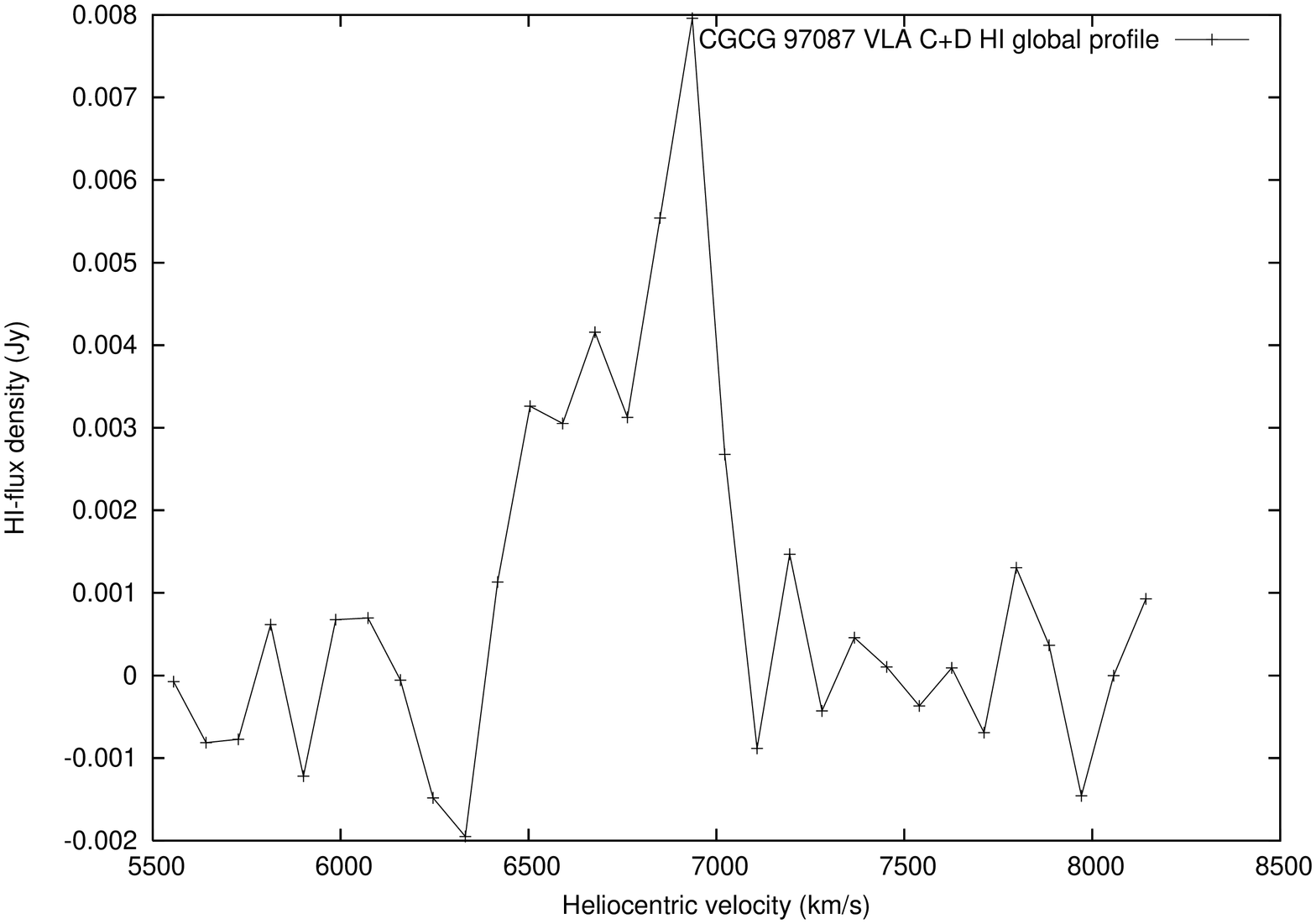}}
\caption{CGCG 097087: H{\sc i}-global profile of the galaxy observed with a spatial and spectral
resolution of $\sim$21$^{\prime\prime}$ and 86 km s$^{-1}$ respectively.
}
\label{87.global}
\end{center}
\end{figure}

The global profiles of H{\sc i} emission made from the combined VLA C-and D-array data with a spectral 
resolution of $\sim$86 km s$^{-1}$ and the VLA D-array data with a spectral resolution of $\sim$22 km s$^{-1}$
are shown in Fig. \ref{79.CD.ispec}, \ref{79.D.ispec} and \ref{79.added.ispec}. The total mass estimated 
from the global spectra is 0.9$\times$10$^9$ M$_\odot$. 
In Fig. \ref{79.CD.chmap} we present the higher spatial resolution channel maps showing the H{\sc i} intensity 
superimposed on the single-channel radio-continuum image shown in gray scale.
The H{\sc i} emission is seen in three channels with velocities of 
7109, 7023 and 6936 km s$^{-1}$ with the emission peak shifting from north-west to
south-east as we go lower in velocities.  DG91 also reported H{\sc i} at two
velocities one having a peak to the north-west with a velocity of 7023 km s$^{-1}$ and the 
other tentative detection on the western boundary of the optical disk at a velocity of 7109 
km s$^{-1}$. We detect emission in a third channel at a velocity of 6936 km s$^{-1}$
giving total velocity width of $\sim$260 km s$^{-1}$.

Moment maps were generated with a 3$\sigma$ cutoff and only three channels with 
a clear detection were combined. A superposition of the the H{\sc i} total-intensity contours
on the DSS blue-band image (Fig. 10) show the south-eastern peak of H{\sc i} emission to be coincident with the
bright starforming region believed to be formed due to the ram pressure 
compression of the ISM. Most of the H{\sc i} emission is on the north-western side of this peak,
along a similar PA as that of the radio-continuum tail. A superposition of the contours of 
H{\sc i} emission on a VLA B-array radio continuum image made from archival data with an angular
resolution of $\sim$4$^{\prime\prime}$ (Fig. \ref{79.CD.M0.radio}) as well as comparison with the continuum image of 
Gavazzi et al. (1995) shows that the radio continuum emission extends well beyond the H{\sc i} emission.  
There is a hint of sharper cut off of the H{\sc i} contours on the western side which is again somewhat
opposite to the arc of star formation, similar to the case of the face on galaxy CGCG 097073.

To examine the velocity field (Fig. 12) with higher spectral resolution we have reduced the VLA D-array data 
which has a spectral resolution of 22 km s$^{-1}$. The channel maps of the image cube with the 
contours of H{\sc i} emission superposed on a single channel
radio continuum image made from the same data using a line-free channel is shown in Fig. \ref{79.D.chmap} .
Significant H{\sc i} emission above a 3$\sigma$ limit has been detected in nine channels with 
velocities ranging from 6923 to 7117 km s$^{-1}$ giving a total width of $\sim$173 km s$^{-1}$.
Because of the relatively poorer spatial resolution of $\sim$45 arcsec the moment 0 image shows
only a blob of emission with no evidence of any diffuse extended emission which may not have been
seen in the higher-resolution image. Hence the moment maps are not presented here. However, to investigate
the velocity field we have fitted  a single Gaussian to H{\sc i} emission in every channel whose position
is marked with a $+$ sign in Fig. \ref{79.skull}. In this Figure the total intensity H{\sc i} emission contours are
superposed on an optical image of the galaxy with the size of the $+$ sign signifying the error in the fit. 
The velocity corresponding to each $+$ sign is indicated in the Figure. The velocity of the extended
emission on the north-western side has a velocity gradient which decreases from $\sim$7120 km s$^{-1}$ 
on its south-western side to $\sim$6920  km s$^{-1}$ on the north-eastern side, with the central velocity
being close to the systemic velocity of the galaxy of $\sim$7000 km s$^{-1}$. The velocity of the H{\sc i} 
gas close to the arc of star formation decreases from $\sim$7050  km s$^{-1}$ on the western side to 
$\sim$6950 km s$^{-1}$ on the eastern side. This shows that the gas in the disk as well as the more extended
extra-planar H{\sc i} gas which has been affected by ram pressure due to the ICM have similar kinematic 
properties with a same sense of rotation. 
It is worth noting that Hota, Saikia \& Irwin (2007) have found similar
properties of the H{\sc i} gas in the edge-on Virgo cluster galaxy NGC4438, where the elongated 
extra-planar gas has a similar sense of rotation as the H{\sc i} gas in the disk of the galaxy.

\subsection{CGCG 097087}

\begin{figure}[h]
\begin{center}
\resizebox{3.5in}{!}
{\includegraphics[angle=0]{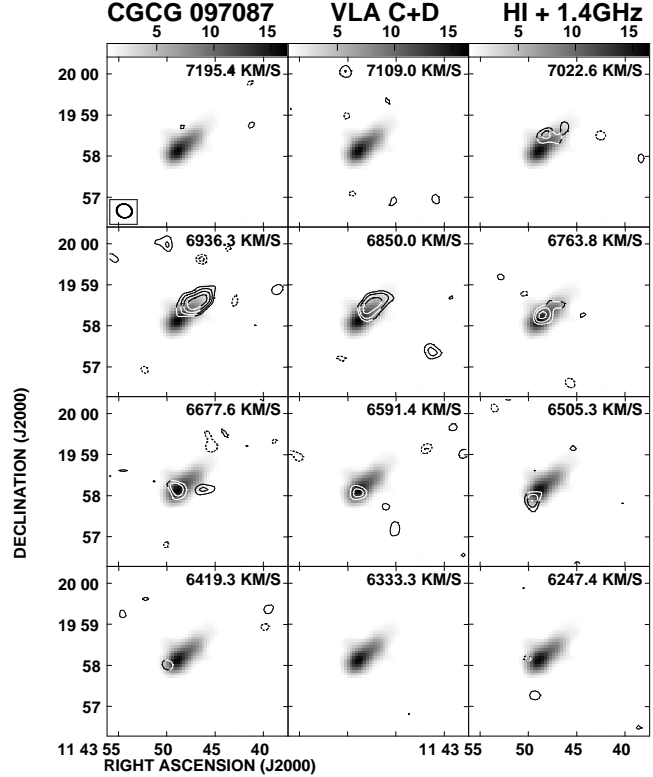}}
\vspace{-1cm}
\caption{CGCG 097087: The H{\sc i} channel maps in contours has been superimposed on the gray scale single 
channel continuum image obtained from the same data with a resolution of $\sim$21$^{\prime\prime}$. 
The contour levels are 0.3$\times$($-$4, $-$2.82, 2.820, 4, 5.65 ...) mJy/beam.
}
\label{87.chmap}
\end{center}
\end{figure}

\begin{figure}[h]
\begin{center}
\resizebox{2.3in}{!}
{\includegraphics[angle=0]{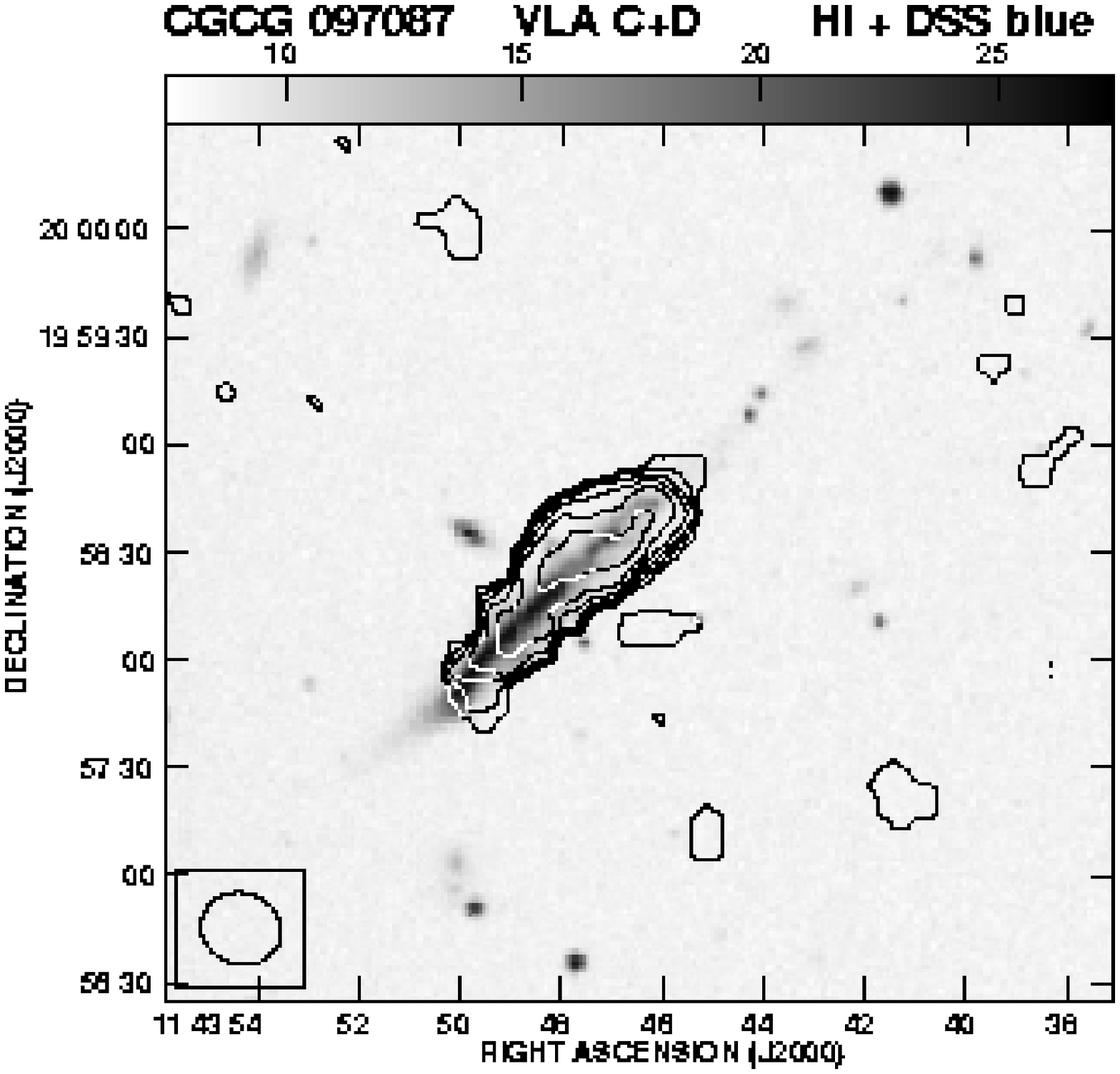}}
\vspace{-1cm}
\caption{CGCG 097087: Total intensity H{\sc i} contour map with a resolution of 
$\sim$21$^{\prime\prime}$ has been superimposed on the DSS blue-band image. The 
contour levels are 2.358$\times$10$^{20}$ atoms cm$^{-2}$ or 1.89 M$_\odot$ pc$^{-2}$ 
and then increasing in steps of $\sqrt{2}$.
}
\label{87.HI.DSS}
\end{center}
\end{figure}

\begin{figure}[h]
\begin{center}
\resizebox{2.3in}{!}
{\includegraphics[angle=0]{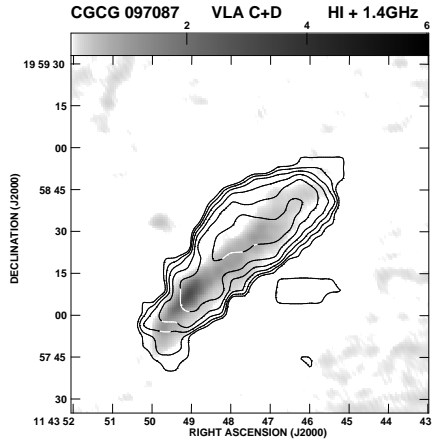}}
\vspace{-1cm}
\caption{CGCG 097087: The same total intensity H{\sc i} contours have been superimposed on the
1.4-GHz radio continuum image made with a higher resolution of $\sim$4$^{\prime\prime}$.
The contour levels are 2.358$\times$10$^{20}$ atoms cm$^{-2}$ or 1.89 M$_\odot$ pc$^{-2}$ 
and then increasing in steps of $\sqrt{2}$.
}
\label{87.HI.radio}
\end{center}
\end{figure}

\begin{figure}[h]
\begin{center}
\resizebox{4.0in}{!}
{\includegraphics[angle=0]{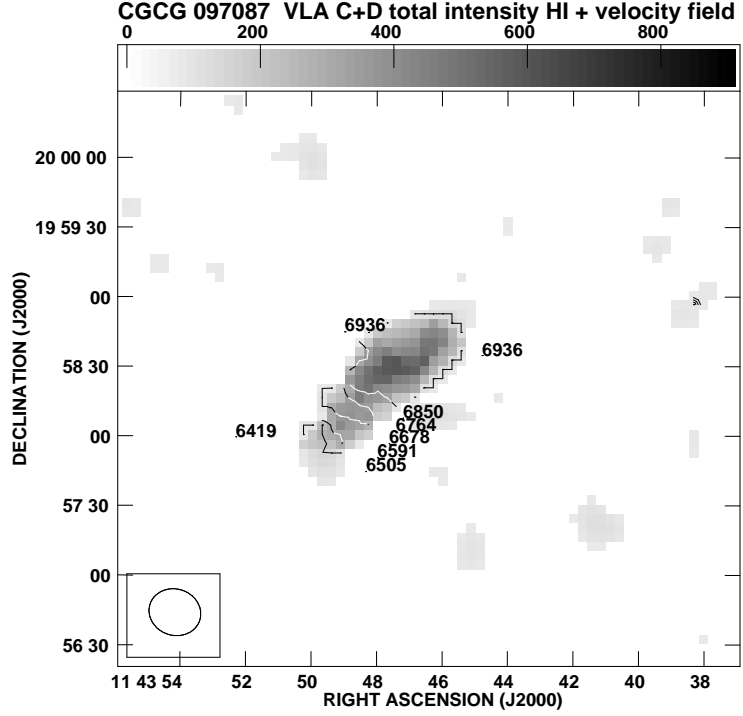}}
\vspace{-1cm}
\caption{CGCG 097087: The intensity weighted H{\sc i}  velocity field has been superimposed on the gray scale image
of the total intensity H{\sc i}. The numbers close to the iso-velocity contours give the heliocentric
velocity at intervals of the channel width of $\sim$86 km s$^{-1}$.
}
\label{87.HIM0M1}
\end{center}
\end{figure}

In Fig. \ref{87.global}  we present the global H{\sc i} profile of CGCG 097087 (UGC 06697) taken from the image cube
which has a spatial and spectral resolution of $\sim$21$^{\prime\prime}$ and 86 km s$^{-1}$ respectively.
The profile has a wide velocity width of $\sim$600 km s$^{-1}$ ranging from $\sim$6435 to 7030 km s$^{-1}$,
and is very much asymmetric with respect to the systemic velocity of 6725 km s$^{-1}$. There is more
gas on the red-shifted side of the systemic velocity, the mass being $\sim$2.9 $\times$ 10$^9$M$_\odot$ 
compared with $\sim$1.7 $\times$ 10$^9$ M$_\odot$ on the blue-shifted side. The spectrum is consistent 
with that obtained with the Arecibo telescope by Gavazzi (1989).  The channel maps of H{\sc i} emission in contours
superposed on a single-channel radio-continuum image
in gray scale from a line free channel with the same resolution is presented in Fig. \ref{87.chmap}.  
The H{\sc i} emission at higher velocities are seen towards the north-west while those with lower
velocities are towards the south-east. DG91 reported detection in three velocity channels, namely at 
6506, 6850 and 6936 km s$^{-1}$. Our recent re-analysis of the data shows emission in 8 channels at 
$\geq$3$\sigma$ level over a velocity range of $\sim$6420 km s$^{-1}$ to 7020 km s$^{-1}$ spanning a
range of $\sim$600 km s$^{-1}$. 

We have generated the moment maps with a cut off at 3$\sigma$ 
and shown the moment 0 image with the total-intensity H{\sc i} emission
contours superposed on the DSS blue-band optical image in Fig. \ref{87.HI.DSS}. 
The total H{\sc i} mass in the region of the bright optical disk of the galaxy 
is $\sim$1.4 $\times$ 10$^9$ M$_\odot$ compared with $\sim$3.2 $\times$ 10$^9$ M$_\odot$ 
for the blob of H{\sc i} emission towards the north-west. The moment 0 image is shown superposed on the
VLA B-array continuum image with an angular resolution of $\sim$4$^{\prime\prime}$
made from archival VLA data (Fig. 18). The tails of radio continuum and H{\sc i} emission are
oriented along very similar PAs and appear to be of similar extents in this Figure.
However, lower resolution images of the tail by Gavazzi \& Jaffe (1987) show that the
tail of radio continuum emission extends to $\sim$3$^{\prime}$ (75 kpc), much larger than the
extent of H{\sc i} emission which extend  only up to 25 kpc from the centre of the galaxy
(defined by the peak of the high resolution radio continuum observation).

The moment 1 map with the iso-velocity contours superposed on the H{\sc i} total-intensity
in gray scale is shown in Fig. \ref{87.HIM0M1}. The H{\sc i} emission coincident with the high-brightness
optical disk shows evidence of solid body rotation, while the north-western region has a
similar velocity of $\sim$6940 km s$^{-1}$. 

\section{Discussion and concluding remarks}
H{\sc i} observations of all the three galaxies CGCG 097073, 097079 and 097087 (UGC 06697)
in the cluster Abell 1367 which have been presented here show the detailed distribution
of the gas. Earlier H{\sc i} observations of these galaxies with the VLA C- and D-arrays 
reported by Dickey \&
Gavazzi (1991) showed the location of a few regions of H{\sc i} emission in different
velocity channels. Our analysis of their data as well as more recent VLA D-array data 
with higher spectral resolution have revealed further details of distribution and
kinematics of the H{\sc i} gas. In all the three galaxies the mass of 
H{\sc i} gas is $\sim$20-30\% larger on the down-stream side of the galaxies, showing
that the distribution of the H{\sc i} gas is affected by
ram pressure of the ICM consistent with earlier suggestions (e.g DG91). 

The directions of the tails of non-thermal radio continuum as well as H$\alpha$
emission suggest that CGCG 097073 is moving towards the south while CGCG 097079
is moving towards the south-east. In CGCG 097073 the H{\sc i} gas    
appears to have a sharper gradient of the contours
on the down-stream side roughly opposite to the arc of starformation region which is
possibly caused by compression of gas due to the ram pressure of the ICM. There is
a suggestion of a similar effect on the western side of CGCG 097079. The sharper
gradient in the H{\sc i} contours may be caused by the accumulation of gas in the
down-stream side due to the effects of ram pressure. The total extent of the H{\sc i} gas 
is $\sim$8 kpc, 8 kpc  and 37 kpc for CGCG 097073, 097079 and 097087 respectively, which 
is much smaller than the 
corresponding tails of non-thermal emission which extend for 75, 60 and 75 kpc
respectively in the low-resolution images (Gavazzi \& Jaffe 1987). The H$\alpha$ tails
also extend for distances of 50, 75 and 55 kpc respectively which are also much
larger than the regions of H{\sc i} emission.    

Results of three-dimensional numerical simulations of a spiral galaxy moving through a hot 
intracluster medium with its disk inclined at different angles to the direction of motion
can produce a wide range of observed structurs (e.g. Quilis, Moore \& Bower 2000; Roediger
\& Br\"uggen 2006, and references therein). 
A comparison of the distribution and kinematics of the H{\sc i} gas in CGCG 097079
with the results of these simulations  shows that the gas in the disk of the galaxy
is pushed backwards by the ram pressure of the ICM. The gas exhibits systematic rotation
about the systemic velocity and is possibly still bound to the parent galaxy.

The direction of the radio tail in CGCG 097087 (UGC 06697) also suggests that this edge-on
galaxy is moving towards the south-east. The H{\sc i} observations show that while the gas
associated with the higher brightness region of the optical galaxy exhibits solid-body rotation,
most of the H{\sc i} gas is pushed towards the north-west and has an almost constant velocity
of $\sim$6940 km s$^{-1}$. 

\section*{Acknowledgments}
We thank the referee Dr G. Gavazzi for a very prompt and helpful report, 
Chanda Jog for discussions and K.S. Dwarakanath for his comments on an early version 
of this paper which has helped present the new results more clearly.
VLA is operated by Associated Universities, Inc. under contract with the National Science
Foundation. This research has made use of the NASA/IPAC extragalactic database
(NED) which is operated by the Jet Propulsion Laboratory, Caltech, under
contract with the National Aeronautics and Space Administration.

\label{lastpage}
\end{document}